\documentclass{meeting_report}
\usepackage{epsfig,caption}
\usepackage{color}
\usepackage{bm}
\usepackage{graphicx}
\usepackage{longtable}
\usepackage{amssymb}
\usepackage{rotating}
\usepackage{latexsym}
\usepackage{hyperref}
\usepackage{float}
\usepackage{flushend}

\newcommand{\rmaxs}{\ifmmode{R_{\rm{\sigma}}^{\rm{max}}}\else{$R_{\rm{\sigma}}^{\rm{max}}$}\fi}
\newcommand{\recirc}{\ifmmode{R_{\rm{e,c}}}\else{$R_{\rm{e,c}}$}\fi}
\newcommand{\re}{\ifmmode{R_{\rm{e}}}\else{$R_{\rm{e}}$}\fi}
\newcommand{\ee}{\ifmmode{\epsilon_{\rm{e}}}\else{$\epsilon_{\rm{e}}$}\fi}
\newcommand{\lr}{\ifmmode{\lambda_R}\else{$\lambda_{R}$}\fi}
\newcommand{\lre}{\ifmmode{\lambda_{R_{\rm{e}}}}\else{$\lambda_{R_{\rm{e}}}$}\fi}
\newcommand{\vs}{\ifmmode{V / \sigma}\else{$V / \sigma$}\fi}
\newcommand{\vse}{\ifmmode{(V / \sigma)_{\rm{e}}}\else{$(V / \sigma)_{\rm{e}}$}\fi}
\newcommand{\vobs}{\ifmmode{V_{\rm{obs}}}\else{$V_{\rm{obs}}$}\fi}
\newcommand{\sobs}{\ifmmode{\sigma_{\rm{obs}}}\else{$\sigma_{\rm{obs}}$}\fi}

\newcommand{\kms}{\ifmmode{\,\rm{km}\, \rm{s}^{-1}}\else{$\,$km$\,$s$^{-1}$}\fi}
\newcommand{\msun}{\ifmmode{M_{\odot}}\else{$M_{\odot}$}\fi}

\newcommand{\mstar}{\ifmmode{M_{\star}}\else{$M_{\star}$}\fi}
\newcommand{\logm}{\ifmmode{\log(M_{\star}/M_{\odot})}\else{$\log(M_{\star}/M_{\odot})$}\fi}

\title{Linking the Galactic and Extragalactic \\ \Huge{A Virtual Meeting During a World-Wide Pandemic}}

\author{Jesse van de Sande$^{1,2}$ and Nicholas Scott$^{1,2}$}

\begin{document}

\maketitle

\begin{abstract}
\large{How do we bridge the gap between the Galactic and the extragalactic? By focusing on the topic of stellar dynamics and stellar populations of the Milky Way and its siblings this virtual meeting aimed at connecting both fields that each bring unique perspectives to understanding how disk galaxies form and evolve. As this meeting took place during a global pandemic, we also give our perspective on the challenges and best practises for running a virtual meeting.}
%
{\let\thefootnote\relax\footnote{
\begin{affiliations}
\item {Sydney Institute for Astronomy, School of Physics, A28, The University of Sydney, NSW, 2006, Australia} 
\item {ARC Centre of Excellence for All Sky Astrophysics in 3 Dimensions (ASTRO 3D), Australia}
\end{affiliations}
}}
\end{abstract}
\vspace{0.5cm}

\noindent Our Milky Way is by far the best studied galaxy in the Universe, unparalleled in resolved stellar measurements of all spectral types. In the last few years, there have been major leaps in our understanding of the Galaxy enabled by the GAIA astrometric mission complemented by many other spectroscopic surveys. With over 7 million radial velocity measurements available from the latest GAIA data release, these data have sparked a detailed investigation into the formation history of the Milky Way from a structural, chemical, and dynamical perspective. 

At the same time, both extragalactic observations and simulations are reaching spatial scales of 10-100 parsec, now allowing for meaningful comparisons to be made between the Milky Way and its extragalactic analogues. The recent wealth of detailed Milky Way measurements from GALAH, APOGEE, LAMOST, and GAIA, combined with results from spatially resolved spectroscopic observations from VLT­-MUSE and large surveys such as SAMI and MaNGA make this the ideal time to link Galactic and extragalactic research. 

The goal of this virtual meeting was to link Galactic and extragalactic research, with their different perspectives on how disk galaxies formed. For a long time our Galaxy has been regarded as a benchmark for understanding the formation and evolution of disk galaxies. Its typical stellar mass, morphology, and colour, make it the ideal template for providing insight into the detailed processes that form and build-up galaxies. Our main aims for this meeting were therefore to: 1) directly compare measurement in the Milky Way with extragalactic observations and state of the art simulations, 2) build connections between the Galactic and Extragalactic community, and 3) challenge the long-held assumption that the Milky Way is a benchmark galaxy. Crucial in this discussion are the recent results from large cosmological and Milky Way zoom-in simulations that shows that the Galaxy might not be the ideal template for understanding disk formation as previously thought.

We divided the meeting into four topics exploring the stellar dynamics and populations of the Milky Way and the stellar dynamics and populations of extragalactic galaxies, as well as an extra topic dedicated to comparisons between the different disciplines. We will give an overview of the key science from this meeting in Section \ref{sec:summary}, whereas the challenges and lessons learnt of transforming an in-person conference into a virtual meeting will be discussed in Section \ref{sec:perspective}.

\section{Summary of the Virtual Meeting}
\label{sec:summary}

\noindent Four key science themes emerged from the meeting: the complexity of mass assembly and disk formation from the Milky Way's perspective, the value (and challenge) of identifying so-called Milky Way analogues, the importance of simulations and analytical models in interpreting between regimes, and the key role recent and future technological innovations have and will play in bringing the two regimes together. \\

\subsection{The complexity of mass assembly and disk formation.} 
Several talks focussed on observations from GAIA and how these have revolutionised our view of the Milky Way. Combined with measurements from large spectroscopic surveys, GAIA has enabled a detailed stellar cartography of the Sun's neighbourhood, showing how the different stellar populations in the Galaxy differ in their kinematics and spatial distributions. Although the chemical structure of the Milky Way is highly variable with location in the disk, broadly speaking the inner Galaxy is dominated by super solar metallicity populations with solar alpha abundances close to the plane, while above the plane the inner Galaxy is dominated by metal poor and $\alpha$-enhanced populations.

Chemical evolution models aim to provide a relatively simple explanation for the origin of the bimodal chemical distribution of Milky Way stars that traditionally have been associated with the structural thin and thick disk components. During the meeting two models were highlighted. The first model assumes a two-step infall approach where the thick disk forms first, in a rapid gas accretion followed by a quenching phase, combined with a second accretion episode that gradually builds the thin disk. The second model has a continuous star formation history and a continuous age velocity dispersion relation, where the thick disk arises naturally from the effects of radial migration and fundamental velocity dispersion relations. This second scheme is gaining traction as it does not require a distinct thick disc component emerging from a separate evolutionary path, and predicts that chemically bimodal disks are commonplace suggesting the Milky Way might be a benchmark galaxy after all.

The Milky Way's stellar halo, its satellites and nearest neighbours also provide a key insights into how the Galaxy assembled its mass over time and how satellites might impact galaxies' internal structure. Recent years have also seen a tremendous increase of observed Milky Way satellites and stellar streams, extending out to possible substructures around the LMC. In particular debris structures offer new insights into hierarchical structure formation on small scales. The best example is the Milky Way's interaction with Sagittarius dwarf where various pericentric passages can be linked with periods of enhanced star formation. However, while 6-dimensional phase space information is available for more than 40 of the Milky Way’s satellite galaxies, just four of Andromeda’s 35 known satellites have available phase space information to date. \\

\subsection{Identifying Milky Way analogues.} Perhaps the most straightforward way to connect Galactic and extragalactic research is by studying Milky Way Analogues (MWAs) that was presented in multiple talks. By selecting extragalactic or simulated galaxies based on a set of physical parameters determined for Milky Way (e.g, stellar mass, star formation rate, disk scale length, morphology), MWAs offer a view of what our Galaxy might look like from an external point of view. MWAs also offer a unique opportunity for studying the impact that certain physical parameters might have on the formation and evolution of disks galaxies, while controlling for others.

From an observational perspective, selecting MWAs can be challenging, as the number of MWAs quickly approaches zero when more or stricter selection criteria are applied. Perhaps naively, one could argue that the Milky Way is therefore unique. Instead it enforces the importance of defining a sample of MWAs suitable for the questions being asked and what science the MWA are used for. Nonetheless, studies selecting MWAs based on having similar structural parameters as our Galaxy, suggest that while the Milky Way has a somewhat low star-formation rate, it is not unusual when compared to its cousins.

Integral Field Spectroscopic (IFS) instruments such as VLT-MUSE and Keck-KCWI, now provide us with the opportunity to obtain high signal-to-noise spectra and unprecedented physical spatial resolution, probing scales of 10-100 parsec. VLT-MUSE observations of a few dozen galaxies with similar mass and morphology as our MW reveal that the central molecular zone of the Milky Way is not a unique structure, with similar components frequently observed in other galaxies and likely created via bar-driven secular processes. Within bar structures, younger stars are trapped on more elongated orbits, forming a thinner component of the bar, whereas older stars form a thicker and rounder component. In general, bars are found to be older, more metal-rich, and less [$\alpha$/Fe]-enhanced than their surrounding disk. While bars are sometimes regarded as transient structures, some bars appear to have formed 10 billion years ago, surviving until the present-day. \\

\subsection{Bridging the Galactic and extragalactic regimes using simulations.} It was also exciting to hear about how modern cosmological simulations and semi-analytic models now provide a large range of possibilities for studying MWAs. The success of such an approach has been demonstrated by several large simulations (e.g. Auriga, EAGLE, FIRE, IllustrisTNG, New-Horizon, NIHAO-UHD, VINTERGATAN), in particular for trying to understand the origin of the thick disk in the Galaxy.

In the New-Horizon simulations the properties of thin and thick discs in galaxies appears to be consistent with observations, but no distinct channels that separate the formation of thin and thick disks are found. These findings are consistent with detailed zoom-in simulations from the NIHAO-UHD, and the thick and thin disk appear to be a result of a continuous in-situ star formation history with a minor accretion contribution. Both results suggest that chemical bimodallity is a common feature of disk galaxies. This is in contrast with results from EAGLE and Auriga. In these simulations only $\sim$5 per cent of MWAs display an [$\alpha$/Fe] bimodality or have chemodynamical properties similar to those of the Milky Way. The lack of chemical bimodality might be connected to its mass-assembly history. These conflicting results from various simulations provide a compelling case to search for a bimodal chemical disk signature in external galaxies. Nonetheless, realistic thin stellar disks traditionally have been difficult to simulate, and remain problematic in large cosmological simulations where individual stellar particle masses are comparable to globular clusters in the Milk Way, casting doubt on the validity of predictions from these simulations.

\subsection{Future prospects.} With all these new measurement also comes an opportunity to employ powerful tools such as Schwarzschild orbit-superposition dynamical modelling. By chemically tagging stellar orbits, spatially resolved maps of the mean stellar population age and metallicity can be linked to different dynamical components and their assembly history. Publicly available codes such as DYNAMITE, now allow us to better exploit high-quality IFS datasets of nearby galaxies. This technique offers a new perspective on connecting chemo-dynamical observations to galaxies' formation histories and are highly complementary to methods using pure stellar light decompositions. \\

\noindent What became clear from trying to connect all talks in this meeting, is that despite the wealth of data from our own Galaxy, the origin of its different components – the thin and thick disks, the bulge, and bar – remain elusive. Our current gap in knowledge was highlighted by the stark contrast between the two closest massive galaxies, the Milky Way and Andromeda. Despite having similar stellar mass and residing in a similar environment, these two galaxies have very different disk, bar, bulge, and halo properties. There is growing evidence that Andromeda's formation history set its apart from the Milky Way, calling the Galaxy its status as a template into question. The dichotomy between these two galaxies and our inability to explain their origin highlights the importance of expanding our research beyond the Local Group. Only by getting a complete census of nearby disk galaxies and their detailed characteristics will allow us to fully understand the formation of different galactic components.


\section{A Perspective on Running a Virtual Meeting}
\label{sec:perspective}

\noindent This meeting was originally planned to be a highly-collaborative, close-knit, in-person event held just outside Sydney, Australia. However the events of early 2020 forced us to dramatically revise our plans, and more importantly our goals for the meeting. Adapting meetings during a time of a worldwide pandemic has proven to be challenging, with a variety of success. The difficulties arising in running virtual meetings are typically not due to insufficiently advanced technology, but from a loss of social interaction and after hours off-topic discussions that often form the foundation of new collaborations and projects. It is currently hard to imagine that online-only meetings, with participants joining from all time zones, will be able to replace all aspects of pre-pandemic in-person meetings (for a detailed discussion see Moss et al. 2021 \cite{moss2021}).

Nonetheless, there are solutions to minimise the strain whilst maximising the value for all participants. For our virtual meeting we adopted a combination of pre-recorded contributed talks (released in the week of the online meeting), a Slack forum for questions and discussions about the pre-recorded talks, and only two days with live sessions including invited talks and discussion sessions led by members of the science organising committee (SOC). Contributed speakers were selected from submitted abstracts using a double anonymous process, and we achieved a fully gender-balanced ratio for all pre-recorded talks. To accommodate various international time zones, the online meeting consisted of two sessions a day: one at 8-10am and one at 8-10pm Australian Eastern Daylight Time \footnote{We note that something as simple as adding a clock to our website featuring the local time, was highly appreciated by many participants and prevented speakers missing their virtual live sessions. To us, this demonstrated that removing as many hurdles as possible is key for running a successful online meeting.} .

While 9am-5pm meetings are already challenging for in-person participation, in a virtual environment with many more distractions available, in particular for participants joining from a non-optimal time zone (e.g., 9pm-5am), it becomes nearly impossible to stay focussed for the entire duration of the meeting. To alleviate this problem we opted for all contributed talks to be pre-recorded. Many options for making videos with slide sharing features now exists (e.g., Zoom, Powerpoint, Keynote), as well as free and reliable online-hosting for all recordings (e.g., Youtube, Vimeo). The many benefits of pre-recorded talks include: 1) access to all talks for all participants at any time from any place, 2) little to no bias arising from talks being scheduled first or last, 3) the ability for each participant to adjust and process information at a suitable pace, 4) as well being able to pause or replay specific talks multiple times. Another advantage of this format is its low-budget costs, with the largest investment being the design and construction of an easy to navigate virtual environment. The Slack channel was integral in enabling discussions between contributed speakers and other participants, which was reflected the large amount of messages ($\sim$2100) sent in a 7 day period. Experiencing this new format with pre-recorded talks was regarded by many as one of the key successes of this meeting. We see a high potential for adopting pre-recorded talks in future \emph{in-person} meeting.

The live sessions consisted of two to three invited talks followed by one hour discussion sessions spanning the four main topics. Choosing the most suitable video communication format was based on: accessibility, the ability to record and produce moderate to high quality videos, and easy to moderate. For this meeting we used Zoom, which is now one of the most commonly used virtual meeting tools in science. It offers a "standard" meeting format as well as a "video Webinar". While the standard option offers the same experience to all participants it also becomes harder to moderate when the number of participants is large. The Webinar format on the other hand, separates panellists from attendees but has the downside that audience members can experience feelings of isolation as direct participation is more restricted and impersonal. As each session had a large set of panellist for the discussion ($\sim$20) combined with an expected audience ranging from 80-120, we adopted the Webinar format.

During the live sessions our choice of running the meeting in the Webinar format became justified. Even though the discussions were led by two members of the SOC, a seemingly simple task of moderating talks, devoting attention to enabling audience engagement via the Q\&A feature to gather audience questions, and encouraging discussion leaders to evenly and fairly spotlight invited and contributed speakers, felt overwhelming. Similar feedback was given by the SOC members who reported that moderating a discussion in the virtual format was significantly harder than any in-person meeting they had attended. Nonetheless, the overall feedback on the meeting was incredibly positive.
 
From our experience running a virtual meeting we believe that relatively small in-person meetings will remain crucial in the foreseeable future to stimulate discussions and promote collaboration, while simultaneously advocating pre-recorded talks for becoming the standard in any format meeting. The latter will spearhead science into becoming a more inclusive and supportive environment for all participants, irrespective of their career-level, race, gender, or background. 


\bibliography{mw_extragal_meeting_report} 

\begin{addendum}
 \item[Acknowledgements] This meeting was made possible through funds from the Australian Research Council Centre of Excellence for All Sky Astrophysics in 3 Dimensions (ASTRO-3D), through project number CE170100013. J.v.d.S. \& N.S. are immensely grateful for the help and contributions from the science organising committee members: James Binney, Joss Bland-Hawthorn, Ken Freeman, Amanda Karakas, Claudia Lagos, Marie Martig, Richard McDermid, and Ricardo Schiavon. We would also like to thank Marie Partridge, Kylie Williams and Kate Gunn, and ASTRO-3D for their help with organising this conference.
\item[Correspondence] Correspondence and requests for materials should be addressed to J.v.d.S. ~(email: jesse.vandesande@sydney.edu.au).
 
\item[Author Contributions] J.v.d.S. \& N.S. were the lead organisers for this meeting and wrote the core text.
\end{addendum}

\end{document}